\documentclass{aa}
\usepackage{times}

\newcommand{\ltsima} {$\; \buildrel < \over \sim \;$}
\newcommand{\simlt}  {\lower.5ex\hbox{\ltsima}}            
\newcommand{\gtsima} {$\; \buildrel > \over \sim \;$}
\newcommand{\simgt}  {\lower.5ex\hbox{\gtsima}}            
\newcommand{\ferg}{erg cm$^{-2}$ s$^{-1}$ }

\newcommand{\be} {\begin{equation}}

\newcommand{\ee} {\end{equation}}
\newcommand{\etal}{{\it et al. }}

\newcommand{\BSAX}{{\em Beppo}SAX} 

\newcommand{\bc}{\begin{center}}
\newcommand{\ec}{\end{center}}

\def \hcm {\hbox {\ifmmode $ atoms cm$^{-2}\else atoms cm$^{-2}$\fi}}

\def\sgr  {SGR~1806--20~}
\def\pdot {\dot P}

\begin{document}
 
 
\title
{  \BSAX\ observations of \sgr }
 
\author{ S.~Mereghetti\inst{1}, D.~Cremonesi\inst{1}, M.~Feroci\inst{2}, M.~Tavani\inst{1} }

\institute {
{1) Istituto di Fisica Cosmica ``G.Occhialini'',
via Bassini 15, I-20133 Milano, Italy\\
2) Istituto di Astrofisica Spaziale, CNR,
via Fosso del Cavaliere 100, I-00133 Roma, Italy}  
}
 
\offprints{S.Mereghetti, sandro@ifctr.mi.cnr.it}
 
\date{Received May 6, 2000 / Accepted June 16, 2000}
 
\authorrunning{S.Mereghetti et al. }
\titlerunning{ \sgr}
\maketitle
 
\begin{abstract}
 
We have observed with the BeppoSAX satellite the quiescent counterpart
of the Soft Gamma-ray Repeater \sgr. Observations performed in October
1998 and in March 1999 showed that this pulsar continued its long term
spin-down trend at an average rate of $\sim$8 10$^{-11}$ s s$^{-1}$ while its
flux and spectrum remained remarkably constant
between the two observations, despite the soft
gamma-ray bursting activity that  occurred in this period.
We also reanalyzed archival ASCA data, that when compared with
the new BeppoSAX observations,  show evidence for a long
term variation in luminosity. 
  
 
\end{abstract}

\section{Introduction}

Soft gamma-ray repeaters (SGRs) are remarkable transient events
characterized by brief ($<$ 1 s) and relatively soft (peak photon energy
$\sim$20-30 keV) bursts of super-Eddington luminosity.
Only four (or possibly five) SGRs are currently known
(see, e.g., Hurley 2000). 
Soon after the discovery of the first three SGRs, the fact that they
were  all located within, or very close to, young supernova remnants
(Vasisht et al. 1994, Kulkarni \& Frail 1993),
and the 8 s periodicity observed during
the famous 1979 March 5 event from SGR~0525--66 (Mazets et al. 1979),
suggested that SGRs  involve some form of   impulsive release of 
energy from a  neutron star. 
This was recently confirmed with the
discovery of periodicities in the 5-8 s range
in the persistent X-ray counterparts of \sgr and SGR 1900+14 
(Kouveliotou et al. 1998, Hurley et al. 1999a), as well as 
in the bursting emission from     SGR 1900+14  
(Cline, Mazets \& Golenetskii 1998). 

 The quiescent 
X--ray counterpart of \sgr was discovered with the ASCA satellite (Murakami et al. 1994)
within  the radio nebula G10.0--0.3, most likely a supernova
remnant (Kulkarni \& Frail 1993). 
Until recently, \sgr  was thought to be associated to a 
luminous blue variable (LBV) star located in the non-thermal core
of the radio nebula (van Kerkwijk et al. 1995). 
Flux variability and    a 
change in the   nebular morphology on a timescale of a few months,
were also detected from the  central part of the radio nebula and thought to be related
to the activity from   \sgr  (Frail et al. 1997).
However, a more precise localization of  this SGR 
(Hurley et al. 1999b) has shown that its
position is incompatible with the LBV, which is probably responsible for 
the non-thermal radio emission, but may be unrelated to \sgr .

Previous X--ray observations of \sgr in the 1-10 keV energy range were obtained with
the ASCA satellite by Sonobe et al. (1994). These authors reported a 2-10 keV
luminosity of  $\sim$10$^{35}$ erg s$^{-1}$  
(for d=10 kpc) and   the following
spectral parameters:  photon index $\alpha_{ph}$ = 2.2$\pm$0.2, 
N$_H$ = (6$\pm$0.2)$\times$10$^{22}$ cm$^{-2}$  and kT = 6.5$\pm$1.5 keV, 
N$_H$ = (5.2$\pm$0.2)$\times$10$^{22}$ cm$^{-2}$
in the case, respectively,  of
a power law and of a  thermal bremsstrahlung fit.

\begin{table*}[htbp]
\begin{center}
  \caption{Summary of the  BeppoSAX observations of \sgr}
  
    \begin{tabular}[c]{ccccc}
\hline
Obs.& Start/Stop                    &  Exposure time$^a$ (ks)     & Count rate$^b$    & Period      \\ 
    &         UT                    & MECS/LECS                   & (counts s$^{-1}$) &  (s)        \\
\hline
A  & 1998 Oct 16 12:30  / 17 10:38  & 30.8 / 10.7        & 0.102$\pm$0.002 & 7.48175 $\pm$ 0.00016 \\
B  & 1999 Mar 21  4:37  / 22 16:26  & 56.7 / 18.2        & 0.103$\pm$0.001 & 7.48271 $\pm$ 0.00003   \\
\hline
 
\end{tabular}
\end{center}
\begin{small}
$^{a}${Net exposure time in the MECS instrument.}\\
$^{b}${Net count rate in 2 MECS units.}\\
\end{small}
\end{table*}

\section{  Observations and Data Analysis  }

The BeppoSAX observations of \sgr were  performed in October 1998 and
in March 1999 (see Table 1 for details). 
Most of our results were obtained with the  Medium  Energy Concentrator 
Spectrometers (MECS) instrument  that covers the  $\sim$ 1.6--10 keV energy range. 

The MECS (Boella et al. 1997)  is based on three position--sensitive 
gas--scintillation proportional counters placed in 
the focal plane of grazing incidence telescopes,  providing images 
over a nearly circular field of view with  $\sim28'$ radius.
The MECS is characterized by
a total (three telescopes) effective area of  $\sim$150~cm$^2$ at 6 keV,
a relatively good angular resolution
(50\% power radius of  $\sim75''$ at 6 keV, on-axis) and 
a moderate energy resolution (FWHM $\sim$8.5$\sqrt{6/{\rm E_{keV}}}$\%).
Due to a failure in  one of the MECS units, only two of them were available
at the time of our observations.

The target was clearly detected on axis in both observations, with a net
count rate in the two MECS units of $\sim$0.1 counts s$^{-1}$.

\subsection{Spectral Analysis}

We  describe the spectral results obtained in the
1.6-10 keV range by using only the
MECS instrument. 
We verified that the inclusion of the data from the LECS instrument 
(Parmar et al. 1997), that extends the energy range down to $\sim$0.1 keV,
did not significantly change the results of the fits. 
In fact the combination of its smaller effective area, shorter exposure times
(see Table 1) and high absorption   gives a small statistics in the
LECS data for \sgr.   
For both observations, the MECS   counts for the spectral analysis were extracted from a circle
of 4$'$ centered at the source position and the background spectra
were estimated from a concentric circular corona with radii $\sim$5$'$ and
10$'$.
In fact the observations of standard  background fields are not adequate in this case
due to the presence of a 
substantial diffuse emission from  the Galactic plane.

The spectra were   equally well  fit with a power law 
($\alpha_{ph}\sim$2) or with a thermal bremsstrahlung (kT$\sim$11 keV)
absorbed by the interstellar
medium (N$_H$  $\sim$6$\times$10$^{22}$ cm$^{-2}$).
A blackbody spectrum  
gave unacceptable results (Obs A: $\chi^2$ = 1.7 / 88 dof,
Obs B:  $\chi^2$ =1.26  / 132 dof).  
Since the results of both observations were consistent with the same spectral
shape and   flux, we also performed a joint analysis of both observations
to further restrict the uncertainties in the spectral parameters under the assumption
of a constant source. 
All the derived parameters are reported in Table 2.

The observed flux for the power law best fit is  
$\sim$ 10$^{-11}$ \ferg in the 2-10 keV energy range.
In the following we assume a distance of 15 kpc, based on
the work by Corbel et al. (1997).
After  correcting for the effect of the 
interstellar absorption,  this corresponds to an
emitted  luminosity of  $\sim$4$\times$10$^{35}$ erg s$^{-1}$ 
(2-10 keV).

\begin{table*}[htbp]
\begin{center}
  \caption{Summary of the MECS Spectral Results}
  
    \begin{tabular}[c]{cccccc}
\hline
Obs.   & Pow Law      &  Absorption    & $\chi^{2}$/dof & 2-10 keV Flux    & 2-10 keV   Flux       \\ 
       & photon index &  10$^{22}$ cm$^{-2}$      &            &     absorbed     &   unabsorbed (\ferg ) \\
\hline
A      & 1.91 $\pm$0.17 &  5.9$\pm$0.8   & 1.31 / 88  &   1.03 10$^{-11}$     &   1.56 10$^{-11}$   \\
B      & 1.98 $\pm$0.12 &  6.5$\pm$0.6   & 0.76 / 132 &   1.06 10$^{-11}$     &   1.68 10$^{-11}$   \\
A+B    & 1.95 $\pm$0.10 &  6.3$\pm$0.5   & 0.98 / 223 &   1.045 10$^{-11}$    &   1.64 10$^{-11}$  \\
\hline
Obs.   &  Bremmstrahlung   &  Absorption    & $\chi^{2}$/dof & 2-10 keV Flux    & 2-10 keV   Flux       \\ 
       & Temperature (keV) &  10$^{22}$ cm$^{-2}$   &            &     absorbed     &   unabsorbed (\ferg ) \\
\hline
A      &  11.7$^{+4.9}_{-2.8}$  &  5.0$\pm$0.6    & 1.34 / 88  &   1.02 10$^{-11}$ & 1.44 10$^{-11}$ \\
B      &  10.6$^{+2.5}_{-1.8}$  &  5.5$\pm$0.5    & 0.808 / 132 &  1.05 10$^{-11}$ & 1.53 10$^{-11}$  \\
A+B    &  11.0$^{+2.1}_{-1.6}$  &  5.35$\pm$0.35  & 1.02  / 223 &  1.04 10$^{-11}$ & 1.50 10$^{-11}$  \\
\hline

\end{tabular}
\end{center}
\begin{small}
Errors are at the  90\% c.l. for a single interesting parameter.
\end{small}
\end{table*}

\begin{figure}[tb] 
\mbox{} 
\vspace{10.5cm} 
\includegraphics{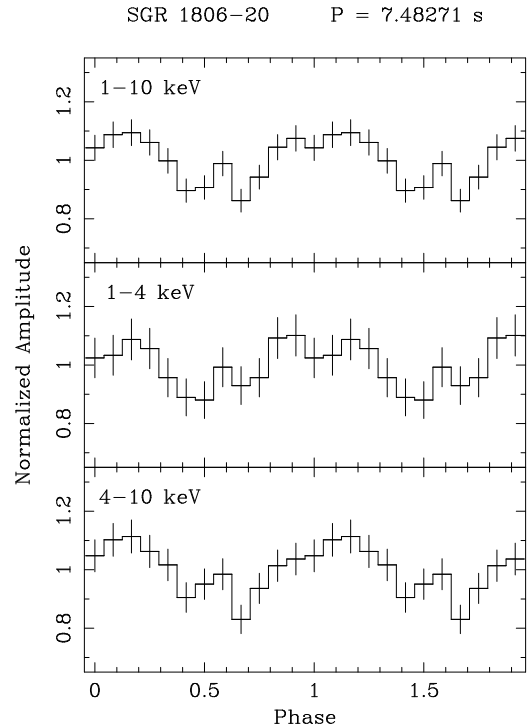} 
 \caption[]{MECS light curves of \sgr folded at the best period for the of the 1999 observation. }
 \label{f2} 
\end{figure}

A comparison   with the previous results reported by Sonobe et al. (1994)
for the October 1993 ASCA observation seems to indicate
that during the BeppoSAX observations the spectrum of \sgr was   slightly
harder and the  luminosity higher. 
Though the evidence for a spectral variation is only marginal and can probably be 
accounted for by 
systematic uncertainties  related to the
comparison between   different instruments, 
the significant  luminosity change (a factor $\sim$2)  deserves a more accurate analysis owing to
its important implications for the nature of \sgr.
We therefore reanalyzed the public ASCA data obtained from the
HEASARC/GSFC on-line archives using the latest available response matrices.

\subsection{Reanalysis of ASCA data}

We analyzed the ASCA GIS data of the Otober 1993 observation.
The data consist of two observations carried out on October 10 and 20.
The spectra were accumulated
from circular regions with radii 6$'$  centered on the source position
and rebinned in order to have at least 25 counts for each energy channel.
As for the BeppoSAX data, we estimated the background spectra from a
source free region in the same observation  to properly take into account the contribution
from the galactic plane X--ray emission.
The power law model gave a very good fit  without evidence
for variations between the two ASCA observations. Combining the two data
sets, we derive the following  best fit values:   $\alpha_{ph}$ = 2.25$\pm$0.15, 
N$_H$ = (6.0$\pm$0.5)$\times$10$^{22}$ cm$^{-2}$,   unabsorbed flux
(1.32$\pm$0.06)$\times$10$^{-11}$ \ferg in the 2-10 keV energy range.
The  values for the spectral index and
absorption are similar to those of Sonobe et al. (1994), but we derive
a slighlty greater flux, closer to the value measured with BeppoSAX.

\subsection{Timing Analysis}

The times of arrival of the  MECS counts used in the spectral analysis were 
converted to the Solar System Barycenter and used in the timing analysis.
The data were first analyzed with a  standard folding technique
and     phase fitting    was subsequently  used to refine the
estimate of the pulse  period. The derived  values are reported in
Table 1. The light curve obtained in the longer observation (1999) is shown 
in Fig.~1 for different energy ranges. 
No significant variations in the shape of the folded light curve are visible
as a function of time and/or energy band.

\begin{figure*}[tb] 
\mbox{} 
\vspace{10.5cm} 
\includegraphics{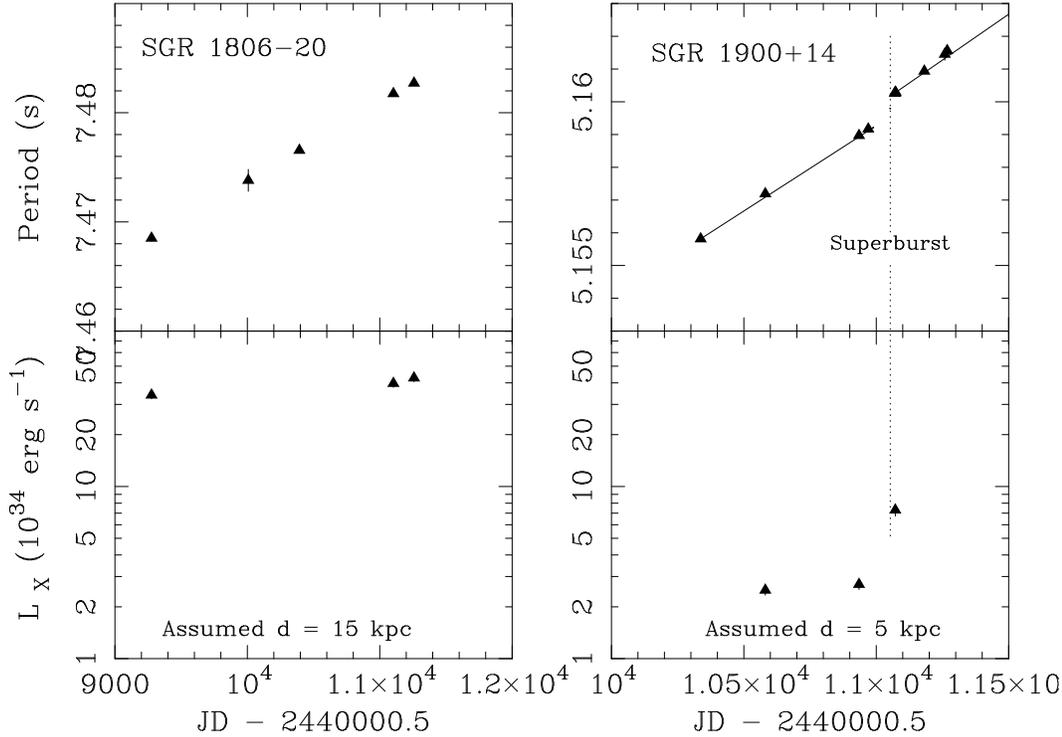} 
 \caption[]{Comparison between   \sgr and SGR~1900+14.The upper panels show the 
long term evolution of the spin period. The vertical dashed line indicates
the time of the 1998 August 27 super outburst of SGR 1900+14.
The unabsorbed 2-10 keV luminosities are indicated in the two bottom panels. 
}
 \label{f2} 
\end{figure*}

\begin{table*}[tbp]
\begin{center}
  \caption{Comparison between \sgr and SGR 1900+14 }
  
    \begin{tabular}[c]{ccccc}
\hline
        & \multicolumn{2}{c} { \sgr }                 &  \multicolumn{2}{c} { SGR 1900+14}   \\ 
\hline

Period & \multicolumn{2}{c} {7.45   s}                         & \multicolumn{2}{c} { 5.16    s}   \\ 
$\pdot$& \multicolumn{2}{c}{8.3$\times$10$^{-11}$ s s$^{-1}$} & \multicolumn{2}{c}{6.1$\times$10$^{-11}$ s s$^{-1}$}\\ 
Distance  & \multicolumn{2}{c} {15 kpc }   &  \multicolumn{2}{c} {5 kpc } \\
 
\hline
                 & ASCA            & BeppoSAX           &   BeppoSAX         &  BeppoSAX     \\ 
                 &  October 1993$^a$   & 1998/99$^a$            &  September 1998$^b$      &  May 1997$^b$  \\
                 &  (Active)      &  (Active)      & (Active)       & (Quiescent)   \\
 \hline
Luminosity   (erg s$^{-1}$)    & 3.4 10$^{35}$  & 4.2 10$^{35}$  & 7.3 10$^{34}$  & 2.5 10$^{34}$ \\
Photon index     &   2.25           &   1.95             &     2.2             &  1.9   \\
Absorption   (10$^{22}$ cm$^{-2}$)    &   6.2            &    6.3             &    2.6              &  1.5   \\
\hline
$^a$ This work\\

$^b$ Woods et al. (1999)
\end{tabular}
\end{center}
\begin{small}
\end{small}
\end{table*}

\section{Discussion}

Soft Gamma-ray Repeaters are usually interpreted in the context
of the so called ''Magnetar'' model  (Duncan \& Thompson 1992; Thompson \& Duncan 1995)
which is based on isolated and strongly magnetized (B$\simgt$10$^{14}$  G) neutron stars.
In this model the magnetic field 
is the main energy source, powering both the  persistent
X--ray (and particle) emission  and the soft gamma-ray bursting
activity. This  involves  internal heating, due to the
magnetic field dissipation,  and the
generation of seismic activity. The latter is responsible 
for the soft $\gamma$-ray  bursts, when the magnetic stresses in the 
neutron star crust shake the magnetosphere
and accelerate particles.

If the spin-down in \sgr is interpreted as purely due
to magnetic dipole radiation losses, the observed values of P and $\pdot$  imply
B$\sim$8 10$^{14}$  G.
Actually, this  value is very likely an overestimate. 
In fact, the particle wind outflow,
either continuous or in the form of strong
episodic outbursts,  also contributes significantly to the spin-down
(Thompson \& Blaes 1998).   Harding et al. (2000) derived the
relations to estimate magnetic field and spin-down age
as a function of the particle wind duty cycle and luminosity. 
They showed that, for instance in \sgr,  a  dipole magnetic filed of 3 10$^{13}$  G  
is derived assuming a continuous particle
wind with a luminosity of $\sim$10$^{37}$ erg s$^{-1}$.

The long term period evolution of \sgr is shown in Fig.~2 (top left panel),
where our results are plotted  with all the previous   measurements
obtained with RXTE and ASCA (Kouveliotou et al. 1998). 
A linear fit to all the points gives $\pdot$=(8.27$\pm$0.18)$\times$10$^{-11}$ s s$^{-1}$.
Woods et al. (2000) reported the results of an extensive
monitoring of \sgr    with the RossiXTE satellite performed from February to August 1999.
The period value we measured in March is consistent with their
timing solution, which however is valid only on a limited time span.
In fact our first period measurement  (October 1998)
is inconsistent with a backward extrapolation of the RossiXTE results.
It is evident that significant variations around the average linear
spin-down trend are present in \sgr. According to Woods et al., the level 
of this ''timing noise'' in \sgr , is relatively larger than that
expected from an extrapolation of that observed in radio pulsars.

When compared with the previous period measurements, our data 
show that, on a long term timescale, the period evolution of \sgr
is relatively stable, despite  the bursting activity that occurred in 1997 and 1998.  
This is in marked contrast  with the behavior of SGR 1900+14 (see Fig.~2, top right panel), 
that displayed  much greater deviations from a constant spin-down rate. 
SGR 1900+14 showed   two distinct periods 
(before and after the Summer of 1998)
of nearly constant spin-down at a similar rate of $\sim$6.1 $\times$10$^{-11}$ s s$^{-1}$. 
The period increase from June to August 1998
could be due either to an enhanced spin-down rate related to the onset
of bursting activity in June 1998 (Woods et al. 1999)
or to a sudden discontinuous jump caused by  the
exceptional ''super burst'' event of August 27  (Feroci  et al. 1999). 
Though plausible explanations for both hypothesis have been advanced
in the context of the magnetar model (Thompson et al. 1999) the lack of
period measurements between June 9 and August 28 does not allow to
discriminate between the two possibilities.   

Although no Soft Gamma-ray Bursts   were detected in our BeppoSAX observations,
we know that \sgr   remained active during the last three years (see, e.g., Woods et al. 2000,
Hurley et al. 1999b). \sgr was also active during the 1993 ASCA observation
that allowed its identification with the steady X--ray counterpart (Murakami et al. 1994).
The lack of large spin-down variations in \sgr indicates that the presence of 
''normal''  bursting activity is not a sufficient condition to generate a 
significant change in the spin-down torques.   This suggests that the
more likely explanation for the spin history of SGR 1900+14 is that
of a discontinuous ''braking glitch'' due to the 27 August 
event. 
 
It is also interesting to compare \sgr and SGR 1900+14 in terms of their
luminosity and long term variability properties (see Table 3 and lower panels
of Fig.~2). 
Our new observations of \sgr  show that this source is extremely stable
in terms of spectral shape,
despite the   evidence of a flux increase with respect to the
1993 level. 
All the observations of \sgr were obtained during periods of bursting activity.
Its luminosity (for d=15 kpc) is about a factor 
$\sim$5-6 higher than that of SGR 1900+14 during its active period of September 1998.
The latter source was even fainter (a factor $\sim$3) 
during observations in May 1997 (BeppoSAX, Woods et al. 1999)
and May 1998 (ASCA, Hurley et al. 1999a), an extended period of quiescence in which
no soft gamma-ray bursting activity was detected.

\section{Conclusions}

New observations of \sgr with the BeppoSAX satellite and
a reanalysis of archive ASCA data have shown that this
source is extremely stable in terms of long term period evolution and spectral 
properties. We   found evidence for a luminosity variation of only 
 $\sim$20\%  between the 1993 value and the more recent 
BeppoSAX observations, both obtained while the source
was in an active state.
This stability, despite the fact that numerous bursts have been observed
in the last three years from \sgr, implies that the normal bursting activity,
independent on its origin, does not significantly affect the
overall (dipolar?) magnetic
field configuration and/or the particle outflow stream  causing the
neutron star spin-down.
Only exceptional events like the 27 August 1998 super outburst of SGR 1900+14
appear to cause significant jumps in the period evolution as well as a
substantially increased X--ray emission.

\end{document}